\documentclass[twocolumn,showpacs,preprintnumbers,amsmath,amssymb]{revtex4}

\usepackage{graphicx}
\usepackage{dcolumn}
\usepackage{bm}
\usepackage{natbib}

\def\yw{Y_{\rm w}}  \def\ew{\eta_{\rm w}}  \def\de{\Delta\eta} \def\bs{B_s}
\def\la{\langle} \def\ra{\rangle}
\def\bs{$\backslash$} 
 
\def\bd{\begin{document}} \def\ed{\end{document}}
\def\bmp{\begin{minipage}} \def\emp{\end{minipage}}
\def\bcc{\begin{center}} \def\ecc{\end{center}}     \def\npg{\newpage}
\def\beq{\begin{equation}} \def\eeq{\end{equation}} \def\hph{\hphantom}
\def\be{\begin{equation}} \def\ee{\end{equation}} \def\r#1{$^{[#1]}$}
\def\n{\noindent} \def\ni{\noindent} \def\pa{\parindent}
\def\hs{\hskip} \def\vs{\vskip} \def\hf{\hfill} \def\ej{\vfill\eject}
\def\cl{\centerline} \def\ob{\obeylines}  \def\ls{\leftskip}
\def\underbar#1{$\setbox0=\hbox{#1} \dp0=1.5pt \mathsurround=0pt
   \underline{\box0}$}   \def\ub{\underbar}    \def\ul{\underline}
\def\f{\left} \def\g{\right} \def\e{{\rm e}} \def\o{\over} \def\d{{\rm d}}
\def\vf{\varphi} \def\pl{\partial} \def\cov{{\rm cov}} \def\ch{{\rm ch}}
\def\la{\langle} \def\ra{\rangle} \def\EE{e$^+$e$^-$} \def\pt{p_{\rm T}}
\def\dt{\delta}
\def\bitz{\begin{itemize}} \def\eitz{\end{itemize}}
\def\btbl{\begin{tabular}} \def\etbl{\end{tabular}}
\def\btbb{\begin{tabbing}} \def\etbb{\end{tabbing}}
\def\beqar{\begin{eqnarray}} \def\eeqar{\end{eqnarray}}
\def\\{\hfill\break} \def\dit{\item{-}} \def\i{\item}
\def\bbb{\begin{thebibliography}{9} \itemsep=-1mm}
\def\ebb{\end{thebibliography}} \def\bb{\bibitem}
\def\bpic{\begin{picture}(260,240)} \def\epic{\end{picture}}
\def\akgt{\noindent{Acknowledgements}}
\def\fgn{\noindent{\bf\large\bf figure captions}}
\def\lan{\langle}
\def\ran{\rangle}
\def\p{\pi}
\def\ifmath#1{\relax\ifmmode #1\else $#1$\fi}%
\def\rc{\ifmath{{\mathrm{c}}}}
\def\cut{\ifmath{{\mathrm{cut}}}}
\def\rF{\ifmath{{\mathrm{F}}}}
\def\rK{\ifmath{{\mathrm{K}}}}
\def\rp{\ifmath{{\mathrm{p}}}}
\def\rt{\ifmath{{\mathrm{t}}}}
\def\LAB{\ifmath{{\mathrm{LAB}}}}
\def\cut{\ifmath{{\mathrm{cut}}}}
\def\beq{\begin{equation}}
\def\eeq{\end{equation}}
\def\us{^{(s)}}  \def\bea{\begin{eqnarray}} \def\eea{\end{eqnarray}}
\def\nbr{\nonumber} \def\e{\eta} \def\D{\Delta}
\def\r{\rho}  \def\unln{\underline}
\newcommand{\cinst}[2]{$^{\mathrm{#1}}$~#2\par}
\newcommand{\crefi}[1]{$^{\mathrm{#1}}$}
\newcommand{\crefii}[2]{$^{\mathrm{#1,#2}}$}
\newcommand{\crefiii}[3]{$^{\mathrm{#1,#2,#3}}$}
\newcommand{\HRule}{\rule{0.5\linewidth}{0.5mm}}
\usepackage{color}
\newcommand{\Blue}[1]{\textcolor[named]{Blue}{#1}}
\newcommand{\blue}[1]{\textcolor[named]{Blue}{#1}}
\newcommand{\Red}[1]{\textcolor[named]{Red}{#1}}
\newcommand{\red}[1]{\textcolor[named]{Red}{#1}}
\newcommand{\violet}[1]{\textcolor[named]{Violet}{#1}}
\newcommand{\brown}[1]{\textcolor[named]{Brown}{#1}}
\newcommand{\green}[1]{\textcolor[named]{Green}{#1}}
\newcommand{\magenta}[1]{\textcolor[named]{Magenta}{#1}}

\def\yw{Y_{\rm w}}  \def\ew{\eta_{\rm w}}  \def\de{\delta\eta} \def\bs{B_s} \def\ssnn{\sqrt {s_{NN}}}
\begin{document}


\title{Statistical and dynamical fluctuations of Binder ratios in heavy ion collisions}

\author{Zhiming Li, Fengbo Xiong, and Yuanfang Wu} \affiliation{Institute
of Particle Physics, Central China Normal University, Wuhan 430079,
China \\ Key Laboratory of Quark and Lepton Physics
(Central China Normal University),Ministry of Education}

\begin{abstract}
Higher moments of net-proton Binder ratio, which is suggested to be
a good observation to locate the QCD critical point, is
measured in relativistic heavy ion collisions. We firstly estimate
the effect of statistical fluctuations of the third and forth order
Binder ratios. Then the dynamical Binder ratio is proposed and
investigated in both transport and statistical models. The energy
dependence of dynamical Binder ratios with different system sizes at
RHIC beam scan energies are presented and discussed.
\end{abstract}

 \maketitle

\section{Introduction}

One of the main goals of current relativistic heavy ion collisions
is to map the QCD phase diagram~\cite{QCDdiagram}. At vanishing
baryon chemical potential $\mu_B = 0$, finite temperature Lattice
QCD calculations predict that a cross-over transition from hadronic
phase to the Quark Gluon Plasma (QGP) phase will occur around a
temperature of 170 - 190 MeV~\cite{Lattice, Crossover}. QCD based
model calculations indicate that the transition could be a first
order at large $\mu_B$~\cite{firstorder}. The point where the first
order phase transition ends is the so-called QCD Critical Point
(QCP)~\cite{QCP1, QCP2}. Attempts are being made to locate the QCP
both experimentally and theoretically~\cite{searchQCP}. Lattice QCD
calculations at finite $\mu_B$ face numerical challenges in
computing. Thus the location of the QCP are highly uncertain in
theoretically. In experimental aspect, the RHIC beam energy scan
program~\cite{RHICbes} has been motivated to search for the QCP in
experiment. By decreasing the collision energy down to a center of
mass of 5 GeV, RHIC will be able to vary the baryon potential from
$\mu_B\sim 0$ to 500 MeV.

Fluctuations of conserved charges, which behave differently between
the hadronic and QGP phase, are generally considered to be sensitive
indicators for the transition~\cite{conserv,fluctuation1}. The
singularity at the QCP, at which the transition is believed to be
second order, may cause enhancement of fluctuations if fireballs
created by heavy ion collisions pass near the critical point during
the time evolution~\cite{fluctuation2}. It has been shown that near
the critical point, the density-density correlator of baryon number
follow the same power law behavior as the correlator of the sigma
field which is associated with the chiral order
parameter~\cite{fluctuation2,baryon}. Therefore, the baryon number
is considered as an equivalent order parameter of formed system in
nuclear collisions. In experiment, net-proton multiplicity
distribution is much easier to measure than the net-baryon numbers.
Theoretical calculations have shown that in QCD with exact isospin
invariance, the relevant corrections due to isospin breaking are
small and the net-proton fluctuations can reflect the singularity of
the baryon number susceptibility as expected at the
QCP~\cite{netproton}. Hence, the net-proton number is used
in current heavy ion experiment~\cite{STARorder}.

It is suggested recently that the Binder-like ratios are good
identification of critical behavior in relativistic heavy ion
collisions\cite{binder,binderwu,binder2}. The
third and forth order Binder ratios are defined as

\begin{eqnarray}
B_{3} &=& \frac{<M^{3}>}{<M^{2}>^{3/2}} \nonumber \\
B_{4} &=& \frac{<M^{4}>}{<M^{2}>^{2}},
\end{eqnarray}

\noindent where $M$ can be a conserved net-charge, e.g. the
net-baryon number. The Binder ratio is a
new observable in heavy ion collisions. The difference between
Binder ratios and the well-known higher
moments\cite{STARorder,highmoment} is that Binder ratios are the
normalized higher raw moments while the
higher moments are central moments.

The universality argument indicates that the static critical
exponents of the second order phase transition are determined by the
dimensionality and symmetry of the system. The QCD critical point of
deconfinement phase transition belongs to the same universality
class as liquid-gas phase transition and the 3D-Ising
model~\cite{fluctuation2,baryon,university}. Its universal critical
properties are discussed to be valid in various of models and
relevant to heavy ion collisions~\cite{fluctuation2, univer1,
univer2}, in particular the event-by-event fluctuations of baryon
numbers~\cite{fluctuation2}.

In the calculations of the 3D-Ising model\cite{binderwu} with
external field $h=0$, Binder
ratios of $B_3$ and $B_4$ as a function of temperature ($T$) show a step
jump from a lower platform to a higher one near the vicinity of
critical point.
If we could map the parameters ($T, h$) of the Ising model onto
the parameters ($T, \mu$) along the freeze-out curve in QCD and find
a path that correspond to $h=0$, the critical behaviors of 3D-Ising model
is expected for the QCD critical point. Therefore, if the formed system
in heavy ion collisions
reaches the critical point and the freeze-out curve is close to the
transition line, the step function liked behavior of Binder ratios
in the Ising model may serve as a probe of QCP in current heavy
ion collisions, where critical incident energy is difficult to
assign precisely in priori.

In the mean time, the effect of trivial statistical
fluctuations~\cite{statistical} due to insufficient number of
particles should be studied and properly eliminated in higher
moment calculations. Therefore, we should discuss the statistical
contributions from the measured fluctuations firstly, then we could
identify the dynamical part which is more relevant to the critical
point of the QCD phase transition.

In this paper, we firstly investigate the statistical and
dynamical fluctuations of net-proton Binder ratios by using the AMPT
and THERMINATOR models. Then, the energy dependence of dynamical
Binder ratios in Au + Au collisions at various RHIC
beam scan energies are studied.

In our analysis, two versions of a multi-phase transport (AMPT)
model~\cite{ampt} are used. One is the AMPT default and the other
one is the AMPT with string melting. In both versions, the initial
conditions are obtained from the heavy ion jet interaction generator
(HIJING) model, and then the scattering among partons is given by
the Zhangs parton cascade (ZPC) model. In the AMPT default model,
the partons recombine with their parent strings when they stop
interacting, and the resulting strings are converted to hadrons
using the Lund string fragmentation model, whereas in the AMPT model
with string melting, quark coalescence is used in combining partons
into hadrons. The dynamics of the hadronic matter is described by
the ART model. The THERMINATOR statistical model~\cite{therminator} is a Monte
Carlo event generator designed for studying of particle production
in relativistic heavy ion collisions from SPS to LHC energies. It
implements thermal models of particle production with single freeze
out.

In order to make our calculations convenient for comparison with the
RHIC beam energy scan data, we choose the mid-rapidity ($|y|<0.5$)
region with transverse momentum $0.4<p_{T}<0.8$ GeV/c. This phase
space is where the STAR experiment can do the particle
identification for proton numbers with its main tracking detector -
the Time Projection Chamber~\cite{STARorder}. The number of events
used in this analysis is around 6 million. This statistics is needed
for the calculation of the dynamical Binder ratios of net-proton to
ensure the statistical errors under control.

\section{Statistical fluctuations of Binder ratios}

\begin{figure}
\includegraphics[scale=0.45]{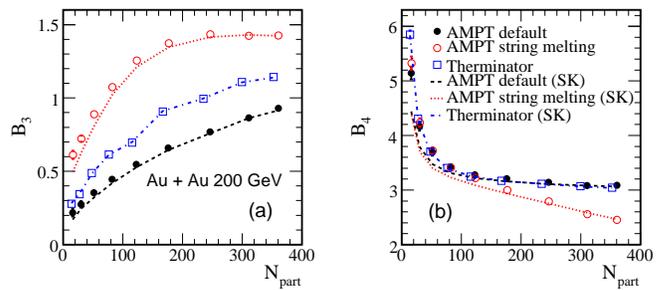}
\caption{\label{fig:epsart} Binder ratios of $B_{3}$ (a) and $B_{4}$ (b)
as a function of number of participants from AMPT default
(solid circle), AMPT with string melting (open circle), and
THERMINATOR (open square) models in Au + Au collisions at
$\sqrt{s_{NN}} = 200$ GeV. The dashed lines represent the corresponding
statistical fluctuations.}
\end{figure}

In the measurement of the net-proton Binder ratios, finite number of
protons and antiprotons will cause non-negligible statistical
fluctuations. If the produced protons and antiprotons are two
independent Poisson-like distributions~\cite{statistical,possion},
the net-protons then obey a Skellam (SK) distribution~\cite{skellam}.

According to the definition of Eq.~(1),  the statistical
fluctuations of the net-proton Binder ratios can be directly deduced
from the Skellam distribution

\begin{eqnarray}
B_{3,stat}&=&
\frac{\Delta^{3}+6\mu\Delta+\Delta}{(\Delta^{2}+2\mu)^{3/2}}\nonumber \\
B_{4,stat}&=&
\frac{\Delta^{4}+12\mu\Delta^{2}+4\Delta^2+12\mu^{2}+2\mu}{(\Delta^{2}+2\mu)^{2}},
\end{eqnarray}

\noindent where $\Delta=\langle N_{p}\rangle-\langle
N_{\bar{p}}\rangle$ is the average number of net-protons, and
$\mu=\left (\langle N_{p}\rangle+\langle N_{\bar{p}}\rangle\right
)/2$ is the mean value of protons and antiprotons in the event sample.
More details of the calculation of this formula could be found in the
appendix.

In Fig.~1 (a) and (b), we show the results of $B_{3}$ and $B_{4}$ as
a function of number of participants ($N_{\rm part}$) from AMPT
default (solid circle), AMPT with string melting (open circle), and
THERMINATOR (open square) models in Au + Au collisions at
 $\sqrt{s_{NN}} = 200$ GeV, respectively. For comparison,
the statistical fluctuations of the Binder ratios, calculated from
Eq.~(2), are presented as dashed lines. We can
see that in both transport and statistical models, the statistical
fluctuations give main contributions to the Binder ratios. It shows
that the influence of statistical fluctuations are not negligible in
the measurement of net-proton Binder ratios at RHIC energy.

\section{Dynamical net-proton Binder ratios}

\begin{figure}
\includegraphics[scale=0.45]{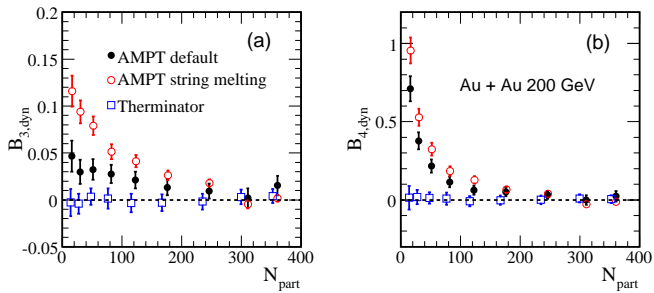}
\caption{\label{fig:epsart} The third (a) and forth (b) order dynamical
Binder ratios as a function of $N_{\rm part}$ from AMPT default
(solid circle), AMPT with string melting (open circle), and
THERMINATOR (open square) models in Au + Au collisions at
$\sqrt{s_{NN}} = 200$ GeV.}
\end{figure}

In the THERMINATOR model, it is well-known that the fluctuations are
thermal. From Fig.~1, we observe it gives a good agreement with the
Skellam statistical fluctuations. It is difficult to disentangle
purely statistical effects from
thermal fluctuations which follow the physics of a hadron resonance
gas. Since neither of them is associate with the QCP behavior, we
suggest to eliminate these statistical or thermal fluctuations in
order to get the dynamical part.

As shown in section II, the statistical fluctuations of Binder ratio
can be expressed by Eq.~(2) given proton and antiproton obey
independent Poisson distributions. We define the so-called dynamical
Binder ratios as,

\begin{eqnarray}
B_{3,dyn}&=& B_{3}-B_{3,stat}\nonumber \\
B_{4,dyn}&=& B_{4}-B_{4,stat}.
\end{eqnarray}

\noindent We suggest to measure these dynamical Binder ratios
instead of the original definition given by Eq.~(1) in relativistic heavy
ion experiment.

The dynamical net-proton $B_{3}$ and $B_{4}$ as a function of
$N_{\rm part}$ from AMPT default (solid circle), AMPT with string
melting (open circle), and THERMINATOR (open square) in Au + Au
collisions at $\sqrt{s_{NN}} = 200$ GeV are shown in Fig~.2 (a) and
(b), respectively. We find that both the third and forth order
Binder ratios from THERMINATOR are zero at all centralities. This is
because that THERMINATOR is based on the hadron resonance gas model
and the produced net-protons in the final state obey the Skellam
distribution~\cite{hrg}.
While, in transport models, both dynamical $B_{3}$ and $B_{4}$ are
larger than zero in peripheral collisions, then tend to be zero in
central collisions. The results from AMPT string melting are larger
than that from the default model. This is due to different
mechanisms of hadronization scheme used for finite state particles
in different versions of AMPT models.

\section{Energy dependence of dynamical Binder ratios in transport models}

\begin{figure}
\includegraphics[scale=0.45]{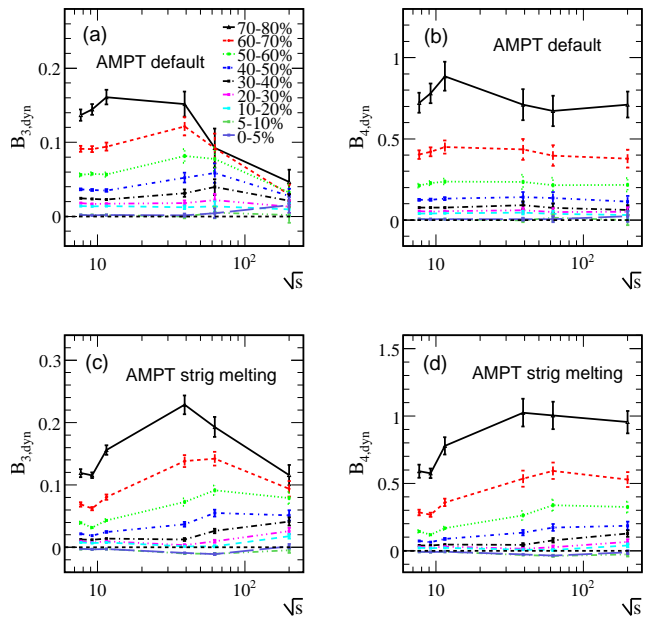}
\caption{\label{fig:epsart} Energy dependence of the
dynamical $B_{3}$ (left panel) and $B_{4}$ (right panel) at RHIC
energies from AMPT default model (upper panel) and string melting (lower
panel), respectively. The nine different symbols represent nine collision sizes, which
goes from most peripheral (70-80\% central) to most central (0-5\% central) collisions.
The lines are used only to guide eyes.}
\end{figure}

In the upper panels of Fig.~3 (a) and (b), we show the energy
dependence of the dynamical $B_{3}$ and $B_{4}$ at six RHIC energies,
7.7, 9.2, 11.5, 39, 62.4, and 200 GeV from AMPT default model.
The nine different symbols represent nine collision sizes (denoted by
centralities in experiments). In all beam energies, when
centrality goes from most peripheral (70-80\% central) to most
central (0-5\% central) collisions, the dynamical Binder ratios
decrease and are close to zero in the most central collisions.
It means that both dynamical $B_{3}$ and
$B_{4}$ are system size dependent. From 7.7 GeV to 200 GeV, we observe no
platform from the AMPT default model. The
lower panels of (c) and (d) for the string melting version give
similar results.

Therefore, there is no step function behavior
observed in both two versions of the AMPT models. This is
understandable since there is no QCD critical mechanism implemented
in these transport models.

\section{Summary and outlook}

In this paper, the statistical and dynamical Binder ratios of
net-proton are studies in Au + Au collisions at RHIC energies. Using
transport and statistical models, it is shown that statistical
fluctuations are not negligible in the measurement of higher Binder
ratios in relativistic heavy ion collisions.

In order to obtain a clean signature which may be related to the
critical point, we suggest to use the dynamical Binder ratio in experimental
measurement. The dynamical net-proton Binder ratio is
found to be zero in the THERMINATOR model but larger than zero in
peripheral collisions in AMPT model. The energy dependence of
dynamical Binder ratios with different system sizes shows no step
function behavior either in AMPT default or string melting models.

Whether the critical behavior of 3D-Ising model without external
filed suggested in ref. [15]
correspond to the QCD critical point could be discussed.
Future work on the study of projecting the QCD parameters onto the
ones in the Ising model and find a path in the phase diagram
that corresponds to the vanishing external field are needed. The alternative way
is that one can include the external field into the Ising model and then
relate it to the Binder ratios to explore the QCD critical point. F. Karsch
et al have explored this way in the 3-state Potts model~\cite{potts}. The
analysis in the O(N) models including the 3D-Ising model is thus called for.

It is interesting to investigate the behavior of the
dynamical Binder ratios in the coming high energy
collisions at RHIC, SPS, and FAIR experiments, where the critical
incident energy of QCD phase transition may be covered. Our model study can
serve as a background study of the behavior expected from known
physics effects for the experimental search for the QCD critical
point.

\section{Acknowledgments}
We thank Dr. Lizhu Chen and Shusu Shi for valuable discussions and
remarks. This work is supported in part by the NSFC of China with
project No. 11005046 and No. 10835005.

\section{Appendix: Binder ratios from Skellam distribution}
Given the distributions of proton and antiproton are independent
Poisson distributions with mean values are $\langle N_{p}\rangle$
and $\langle N_{\bar{p}}\rangle$, the net-proton will follow a
Skellam distribution. If we define the net-proton as
$M=N_{p}-N_{\bar{p}}$, then the probability distribution function of
$M$ is
\begin{eqnarray}
&&f(M;\langle N_{p}\rangle,\langle N_{\bar{p}}\rangle ) \nonumber \\
&=&e^{-(\langle N_{p}\rangle+\langle N_{\bar{p}}\rangle)}\left (
\frac{\langle N_{p}\rangle} {\langle N_{\bar{p}}\rangle}\right
)^{M/2}I_{|M|}\left ( 2\sqrt{\langle N_{p}\rangle \langle
N_{\bar{p}}\rangle} \right ),\nonumber
\end{eqnarray}

\noindent where $I_{|M|}\left ( 2\sqrt{\langle N_{p}\rangle \langle
N_{\bar{p}}\rangle} \right )$  is the modified Bessel function of
the first kind.

The $n^{th}$ moment of $M$, which is defined as $\langle
M^{n}\rangle=\int_{-\infty}^{\infty} M^{n}f(M;\langle
N_{p}\rangle,\langle N_{\bar{p}}\rangle )dM$, can be calculated from
the above distribution function. We obtain
\begin{eqnarray}
\langle
M\rangle&=& \Delta\nonumber \\
\langle
M^{2}\rangle&=& \Delta^{2}+2\mu\nonumber \\
\langle
M^{3}\rangle&=& \Delta^{3}+6\mu\Delta+\Delta\nonumber \\
\langle M^{4}\rangle&=&
\Delta^{4}+12\mu\Delta^{2}+4\Delta^{2}+12\mu^{2}+2\mu,\nonumber
\end{eqnarray}

\noindent where $\Delta=\langle N_{p}\rangle-\langle
N_{\bar{p}}\rangle$ is the average number of net-protons, and
$\mu=\left (\langle N_{p}\rangle+\langle N_{\bar{p}}\rangle\right
)/2$ is the mean value of proton and antiproton.

By the definitions of the third and forth Binder ratios of Eq.~(1),
we get the Binder ratios of the Skellam statistical distribution as

\begin{eqnarray}
B_{3,stat}&=& \frac{<M^{3}>}{<M^{2}>^{3/2}} =
\frac{\Delta^{3}+6\mu\Delta+\Delta}{(\Delta^{2}+2\mu)^{3/2}},\nonumber \\
B_{4,stat}&=& \frac{<M^{4}>}{<M^{2}>^{2}} =
\frac{\Delta^{4}+12\mu\Delta^{2}+4\Delta^2+12\mu^{2}+2\mu}{(\Delta^{2}+2\mu)^{2}}.\nonumber
\end{eqnarray}


\end{document}
%